\documentclass[pre,twocolumn,twoside,aps,floatfix,longbibliography,natbib,nofootinbib,superscriptaddress]{revtex4-1}

\usepackage{amsmath, bm, amsfonts, amssymb}     

\usepackage{graphicx}                       
\usepackage{xfrac}                          
\usepackage{grffile}                        
\usepackage{color,colortbl}
\usepackage{xcolor}

\usepackage[colorlinks, citecolor=blue, urlcolor=red, linkcolor=blue]{hyperref}

\newcommand{\sig}{\sigma_0}
\newcommand{\sigmay}{\sigma_{\rm Y}}
\newcommand{\sigmass}{\sigma_{\rm ss}}
\newcommand{\xg}{x_{\rm g}}
\newcommand{\tstop}{t_{\rm stop}}
\newcommand{\gamf}{\Delta\gamma_{\rm f}}
\newcommand{\gamr}{\Delta\gamma_{\rm rec}}
\newcommand{\tw}{t_{\rm w}}
\newcommand{\gdot}{\dot{\gamma}}

\newcommand{\be}{\begin{equation}}
\newcommand{\ee}{\end{equation}}
\newcommand{\bea}{\begin{eqnarray}}
\newcommand{\eea}{\end{eqnarray}}

\newcommand{\smf}[1] {\textcolor{black}{#1}}
\newcommand{\smfb}[1] {\textcolor{black}{#1}}
\newcommand{\smfc}[1]{\textcolor{black}{#1}}
\newcommand{\smfd}[1]{\textcolor{black}{#1}}

\begin{document}

\title{\smfc{Recoverable strain in amorphous materials: the role of ongoing plastic events following initial elastic recoil}}
\author{Henry A. Lockwood and Suzanne M. Fielding}
\affiliation{Department of Physics, Durham University, Science Laboratories, South Road, Durham DH1 3LE, UK\newline Corresponding author: suzanne.fielding@durham.ac.uk}

\begin{abstract}

Recoverable strain is the strain recovered once a stress is removed from a body, in the direction opposite to that in which the stress had acted. To date, the phenomenon has been understood as being elastic in origin: polymer chains stretched in the direction of an imposed stress will recoil after the stress is removed, for example. Any unrecoverable strain is  instead attributed to irreversible plastic deformations. Here we study theoretically \smfc{strain recovery} within the soft glassy rheology model, aimed at describing the rheology of elastoplastic yield stress fluids and amorphous soft solids. We consider a material subject to the switch-on of a shear stress that is held constant before later being set back to zero, after which the strain recovery is observed.  \smfc{After an initially fast recoil that is indeed elastic in nature, significant further strain recovery then occurs more slowly via the plastic yielding of elements with negative local stresses, opposite to that of the original shear. We elucidate the mechanism that underlies this behaviour, in terms of the evolution of the SGR model's population of elastoplastic elements. In particular, we show that the initial fast elastic recoil brings to a state of negative local stress those elements that had yielded during the forward straining while the load was applied. The subsequent delayed  plastic yielding of these elements with negative stress is the origin of the slow ongoing strain recovery. In this way, counterintuitively,  elements that had yielded plastically while the load was applied still contribute significantly to strain recovery after the sample is unloaded. This finding has important consequences for constitutive modeling, because such behaviour can only arise in a constitutive model that evolves a full distribution of local stresses (or multiple moments of such a distribution), rather than a single average stress.} Unexpectedly, although in rare parameter regimes, this \smfc{slow ongoing} strain recovery post switch-off does not always in fact recover in the negative direction, counter to that of the previously imposed stress, but can sometimes continue to accumulate in the forward direction. The recovery is then non-monotonic overall, reminiscent of observations of non-monotonic stress relaxation after straining. 

\end{abstract}

\maketitle

\section{Introduction}
\label{sec:intro}

The concept of recoverable strain has a long history in the rheology literature. Reiner noted that, when stresses are removed from a body, part of the deformation will in general be recovered, and that this part of the deformation is elastic~\cite{reiner1958rheology}. Weissenberg suggested that the stress in a flowing material can be expressed at any  time in terms of a strain that relates the current state of the material to some reference state; and that this reference state, which depends on the flow history, defines the recoverable state to which the material would ultimately deform if the stress were then removed~\cite{weissenberg1947continuum}. In essence, then, recoverable strain is the amount of strain recovered once a stress is removed from a body, in the direction opposite to that in which the stress had acted. 

In physical terms, recoverable strain has to date been understood as elastic in origin: polymer chains elastically stretched in the direction of an imposed stress show viscoelastic recoil after the stress is removed, for example. (In a material with a single Maxwell relaxation timescale, this recovery is predicted to occur instantaneously. A superposition of Maxwell modes with distinct relaxation timescales confers a finite recovery time~\cite{white1975considerations}.) Indeed, the term `elastic recovery' has often been used synonymously with `strain recovery'~\cite{lodge1958network,mooney1936rheology,reiner1945classification,white1975considerations}. Any unrecoverable strain is instead attributed to irreversible plastic deformations. 

The same distinction is also evident in more recent literature concerning the use of recoverable vs unrecoverable strain to characterise yielding~\cite{donley2020elucidating,shim2023understanding,kamani2021unification,lee2019structure}. Donley et al.~\cite{donley2020elucidating} note that zero-stress recovery tests allow materials to recover their instantaneous `ground state', enabling calculation of the relative amounts of viscosity and elasticity. Shim and Rogers~\cite{shim2023understanding} state that `the recoverable and unrecoverable [strain] components can be interpreted as the contributions from the viscoelastic solid and plastic properties respectively'. Kamani et al.~\cite{kamani2021unification} note that  `the recoverable [strain] component is related to elastic processes, while the unrecoverable component is related to the plastic behaviour' and that `yield stress fluids change from being viscoelastic solids, where deformations are recoverable, to deforming plastically, where deformation is unrecoverable'. Lee et al.~\cite{lee2019structure} state that `recoverable strain is elastic, while viscous properties are dictated by the rate at which strain is acquired unrecoverably'.

In experimental studies, strain recovery following an initial forward creep under an imposed stress, after the stress is subsequently switched-off, has been observed in  cold set gels~\cite{brito2022modeling}, protein gels~\cite{leocmach2014creep}, fractal colloidal gels~\cite{aime2018power}, hectorite clay~\cite{ten2007rheology}, peptide gels~\cite{helen2011mechanosensitive}, carbopol~\cite{lidon2019mesoscale} and polystyrene aggregate concrete~\cite{tang2014creep}. A high degree of recovery after a large deformation has been discussed as being an important property of tough hydrogels~\cite{zhao2014multi,sun2012highly,bakarich2012recovery,hou2017rapidly}. Shape memory effects in self-healing polymers are also attracting increasing attention~\cite{hornat2020shape}, although here the recovery is typically induced by a change in temperature, or other control parameter besides stress or strain, which we do not consider here.

In this work, we study theoretically  recoverable and unrecoverable strain within the class of materials that are variously (in different parts of the literature) referred to as soft glassy materials~\cite{sollich1997rheology,sollich1998rheological}, amorphous materials~\cite{falk2011deformation}, yield stress fluids~\cite{lindeman2021multiple} or elastoplastic materials~\cite{nicolas2018deformation}. These include dense foams and emulsions, glassy colloidal suspensions, jammed athermal suspensions and granular matter, as well as harder metallic and molecular glasses. We do so by performing simulations within the widely used soft glassy rheology model~\cite{sollich1997rheology,sollich1998rheological}, focusing in particular on the simple recovery protocol in which a material of age $\tw$ is subject to the switch-on at some time zero of a shear stress of amplitude $\sig$, which is held constant until it is later  set back to zero at some time $\tstop$, after which the strain recovery is  observed as a function of the time $t-\tstop$. 

\smfc{After an initially fast recoil that is elastic in nature, we find, counterintuitively,  that  significant further slow strain recovery then occurs via ongoing plastic events in the reverse direction, opposite to that of the original shear. We provide a mechanistic understanding of this phenomenon in terms of the evolution of the SGR model's population of elastoplastic elements. In particular, we show that the initial recoil, which elastically advects each element backwards in its trap, leaves in a state of negative stress those elements that had yielded plastically while the material was under load. The subsequent delayed plastic yielding of these elements with negative stress then leads to slow ongoing strain recovery. Because this yielding is of {\em negative stresses}, counter to the original shear, we call these ``reverse plastic events". Counterintuitively, therefore, those elements that had yielded plastically while the forward load was applied still make an important contribution to the material's overall recovery after unloading. This has important consequences for constitutive modeling, because such behaviour can only arise in a constitutive model that evolves a full distribution of local stresses (or multiple moments of such a distribution).  A model that instead evolves only a single average stress will be unable to capture it.  To illustrate this, we also simulate a simple fluidity model, which shares many features of SGR's behaviour, including a yield stress, rheological aging, and stress overshoot in shear startup. In evolving only an average stress rather than a stress distribution, however, it lacks the slow ongoing strain recovery predicted by the SGR model.}

\smfc{We show that the observation of slow ongoing strain recovery from reverse plastic events holds robustly across broad parameter regimes. Unexpectedly -- although in relatively rare parameter regimes -- we also find that the slow ongoing strain post switch-off does not always  take place in the negative direction, counter to that of the originally imposed stress, but can sometimes continue to accumulate in the forward direction. The recovery process is then non-monotonic overall, with the initially backward elastic recoil followed by a (lesser) slow forward straining. This is reminiscent of recent observations of non-monotonic stress relaxation after straining~\cite{hendricks2019nonmonotonic,sudreau2022residual,murphy2020memory,mandal2021memory}, which have been attributed to complex material memory effects~\cite{keim2019memory}. }

\smfd{The notion of a distribution of local stresses in an amorphous material, rather than a single macroscopic average stress, has a well developed history in the statistical physics literature. Modelling approaches based on shear transformation zones~\cite{falk1998dynamics} and lattice elastoplastic descriptions are based upon it~\cite{hebraud1998mode,nicolas2018deformation,bocquet2009kinetic,lin2014scaling}, as is the SGR model itself~\cite{sollich1997rheology,sollich1998rheological}.  These approaches in turn draw on concepts originally developed by Spaepan~\cite{spaepen1977microscopic} and Argon~\cite{argon1979plastic} in the context of metallic glasses and bubbles rafts. More recently, frustrated local stresses have been recognised to play an important role in the elastically driven aging of soft solids~\cite{bouzid2017elastically}, the   memory of soft jammed solids to shear flow~\cite{vinutha2024memory}, the elasticity properties of amorphous solids~\cite{zhang2022prestressed}, the directional memory of soft glasses~\cite{edera2025mechanical}, and the non-monotonic relaxation of shear stress after flow cessation~\cite{ward2025shear}. A key contribution of this work is to recognise and explore the significance of a stress distribution in strain recovery after unloading, and to emphasise that important effects such as non-monotonic recovery can be captured only in constitutive descriptions that indeed consider a full distribution of local stresses (or at least multiple moments of such a distribution), rather than a single average stress.}

The paper is structured as follows. In Sec.~\ref{sec:model} we outline the SGR model~\cite{sollich1997rheology,sollich1998rheological}, within which we shall study the phenomenon of strain recovery. The step stress protocol that we shall simulate is specified in Sec.~\ref{sec:protocol}. In Sec.~\ref{sec:parameters} we define our units,  summarise the parameters of the model and protocol, and discuss  the rheological quantities to be measured. Our results are presented in Sec.~\ref{sec:results}. We set out our conclusions in Sec.~\ref{sec:conclusions}.

\section{Soft glassy rheology model}
\label{sec:model}

The soft glassy rheology (SGR)
model~\cite{sollich1997rheology,sollich1998rheological}
considers the dynamics of an ensemble of elastoplastic elements that explore a landscape of energy traps. Each element is notionally taken to represent a
local mesoscopic region of a soft glassy material: a few tens of
droplets in a dense emulsion, for example. Each element is assumed small relative to any macroscopic variations in the flow field, on the one hand, yet large enough to allow the definition of local continuum
variables of shear strain $l$ and shear stress $kl$, on the other. \smf{Here $k$ is an elastic constant, assumed the same for all elements.} The local strain variable $l$ is taken to describe
the element's state of elastic deformation relative to a locally
undeformed equilibrium.  Between local plastic yielding events, defined below, the strain
of each element is assumed elastically to follow the macroscopically imposed strain rate $\gdot$, such that
$\dot{l}=\gdot$. 

Plasticity is incorporated by assuming that these local stresses are intermittently released by local plastic yielding
events. In each such yielding event, a mesoscopic region of material is assumed suddenly to  rearrange itself
into a new local configuration, relative to a new state of local equilibrium. This is modelled by the corresponding representative
elastoplastic element hopping between traps in the model's energy
landscape: at any instant in time, an element that lies in a trap of energy depth $E$ and has local shear
strain $l$ is assumed to have a probability per unit time of hopping \smf{given by~\cite{sollich2012thermodynamic}}
\be
\label{eqn:rate}
r(E,l)=\tau_0^{-1}{\rm min}\left\{1,\exp\left[(\tfrac{1}{2}kl^2-E)/x\right]\right\},
\ee
with $\tau_0$ a microscopic attempt time.
As can be seen via this expression, the  elastic strain energy $\tfrac{1}{2}kl^2$ essentially reduces  the local energy barrier $E$, giving an effective barrier $E-\tfrac{1}{2}kl^2$. This shear-induced reduction in barrier heights  renders the model's overall rheological behaviour shear thinning. Depending on the material considered, the parameter $x$ can be interpreted as the true thermal temperature or an effective noise temperature: a point to which we shall return below.

After hopping, an element is assumed instantaneously to select a
new trap depth randomly from the landscape's prior distribution of trap depths, taken to be of the form:
\be
\label{eqn:prior}
\rho(E)=\frac{1}{\xg}\exp\left(-E/\xg\right),
\ee
in which $\xg$ is the model's glass transition (effective) temperature, discussed further below.
The element is further assumed to reset its local strain $l$ to zero, so relaxing its
stress. \smf{(At the end of Sec.~\ref{sec:results}, we shall return to consider a distribution of post-hop strain values with non-zero width $l_p$, to model local frustration.)}

With these dynamics, the joint probability $P(E,l,t)$ for an
element to be found at any time $t$ in a trap of depth $E$, and with local shear strain $l$,
evolves via the following master equation:
\be
\label{eqn:master}
\dot{P}(E,l,t)+\gdot\frac{\partial P}{\partial l} = -r(E,l)P+Y(t)\rho(E)\delta(l).
\ee
The advection term on the left hand side  captures the elastic
loading of each element within its trap. The first term on the
right hand side describes the local plastic yielding events described above. Because these are modelled by hops out of traps, this takes the form of a `death' term. The second term on the right hand side is a `rebirth' term,  describing the hopping of freshly yielded elements into the bottom of new traps, $l=0$, with the new trap depth selected randomly from the prior  $\rho(E)$ of Eqn.~\ref{eqn:prior}.  The ensemble average hopping rate
\be
Y(t)=\int dE \int dl\, r(E,l)\,P(E,l,t),
\ee
and the ensemble average macroscopic stress
\be
\label{eqn:sigma}
\sigma(t)=k\int dE \int dl\; l \,P(E,l,t).
\ee

In combination with the exponential prior distribution of trap depths across the model's energy landscape, $\rho(E)$, the exponential activation factor $r(E,l)$ confers a glass 
 transition at an effective temperature $x=\xg$.  In the absence of any imposed shear, the model captures rheological ageing for effective temperatures in the glass phase $x<\xg$~\cite{fielding2000aging}. Following the preparation of a sample 
via a sudden quench from a high initial effective  temperature to a working temperature $x<\xg$, the system slowly evolves into ever deeper traps as a function of the time since the temperature quench. The overall stress relaxation time grows linearly with time, giving progressively more solid-like rheology as the sample ages.

Conversely, an imposed shear of rate
$\gdot$ will halt this ageing process and rejuvenate the sample to a flowing state with an  effective age set by the inverse shear rate, $1/\gdot$. In its glass phase, $x<\xg$, the model predicts  steady state flow curves with a yield stress $\sigmay(x)$ in the limit of slow shear $\gdot\to 0$. For $\xg<x<2\xg$ the model predicts power law flow curves with a shear-thinning (sublinear) exponent, giving infinite viscosity in the limit of slow shear. For  $x>2\xg$, the flow curves are Newtonian.

In many soft glassy materials, the typical scale of energy barriers for local
rearrangement events -- the penalty for stretching soap films in a foam, for example --  is far greater than the typical scale of thermal energy, $k_{\rm B}T$. In such materials, the
parameter $x$ is  taken to be
an effective noise temperature that models an assumed 
coupling with other yielding events elsewhere in the sample. This is intended to capture in a mean-field way the basic idea that a local yielding event in one part of the sample may trigger follow-on yielding events in other regions. To date, this remains a mean-field assumption that is yet to be made self consistent. In materials for which the energy barriers are instead on the scale of $k_{\rm B}T$, the parameter $x$ can be taken as the true thermal temperature.

\section{Protocol}
\label{sec:protocol}

In this section, we define the recovery protocol that we shall simulate within the SGR model just described.

Sample preparation prior to shear is modelled by assuming a sudden deep quench  at some time $t=-\tw$, from infinite effective temperature to the final working value $x$, which is held constant thereafter. For times $-\tw<t<0$ the sample is allowed to age undisturbed, without any strain or stress being applied. Throughout most of the paper, we assume that all elemental strains are equal to zero during this initial ageing stage, $l=0$. \smf{An exception can be found in Figs.~\ref{fig:basic} and~\ref{fig:distributions}, where we assume an initial Gaussian distribution of $l$ values, of narrow standard deviation $l_0=0.05$. This does not affect the basic physics of interest in this work, but merely allows us to show  as a narrow Gaussian in Fig.~\ref{fig:distributions} what would otherwise be an unplottable initial delta function distribution. We return at the end of Sec.~\ref{sec:results} to consider in more detail the effect of non-zero $l_0$ values on the physics that we report.}

At time $t=0$, a step stress of amplitude $\sig$ is imposed. At the time of this step, all elements instantaneously strain elastically forward within their traps by the same amount $l=\sig/k$, giving an instantaneous global elastic strain response $\gamma_0=\sig/k$.  With the imposed stress held constant and equal to $\sig$, we then allow the sample to strain further forward by an amount $\gamf$, beyond the elastic straining that occurred at the instant the stress was imposed. Note that this further straining arises from elements hopping between traps, and accordingly is plastic in nature.  Once the total strain has attained a value $\gamma(t)=\gamma_0+\gamf$, we  set the stress to zero, $\sigma=0$. We define the time at which this occurs as $t=\tstop$. In principle, we could use either $\tstop$ or $\gamf$ as a control parameter in our study. In practice, we have chosen to use $\gamf$ because it better reflects the physical state of the material at switch-off. 

For all subsequent times, $t>\tstop$, we hold the stress equal to zero and measure the strain $\gamma$ as a function of the time interval $t-\tstop$ since switch-off. We shall be interested in particular in the degree to which the strain recovers in the negative direction, whether it does so monotonically or non-monotonically, and in the basic physics that underlies this recovery process. 

\section{Units, parameter values and measured quantities}
\label{sec:parameters}

The parameters of the SGR model are the effective temperature $x$, the attempt time for local yielding events $\tau_0$, the elemental modulus $k$, and the glass transition temperature $\xg$.  Recall Sec.~\ref{sec:model}. The parameters of the rheological protocol  are the time $\tw$ for which the sample is aged before the stress is imposed, the amplitude of the imposed stress $\sig$, and the degree to which the sample is allowed to strain further forwards due to plastic yielding while the stress is held fixed, $\gamf$. Recall Sec.~\ref{sec:protocol}.

\smf{To solve the SGR model numerically, (at least) two different methods can be used. In the first, one derives integral equations from the governing master equation, Eqn.~\ref{eqn:master}, then uses numerical mathematics to solve these~\cite{sollich1998rheological}. In the second method, which we adopt here, one instead directly simulates the stochastic hopping dynamics of a population of $M$ SGR elements. In any timestep $dt$, each element has its strain shifted elastically by an amount $l\to l+\gdot(t) dt$, where $\gdot(t)$ is the strain rate at that time. Any element also plastically hops into a new trap with probability $r dt$, with $r$ that element's probability per unit time of hopping as given by Eqn.~\ref{eqn:rate} above.  In the limit of many elements $M\to\infty$, one recovers the solution of the governing equations. During any intervals in which the stress is held constant, the strain rate is calculated as $\gdot=\langle lr \rangle_P$. This ensures that the rate of stress loss via plastic hopping between traps is exactly counterbalanced by stress gain via elements shifting their local strains within their existing traps. At the instant of any step stress of size $\Delta \sigma$, each element is simply shifted elastically in its trap by an amount $l\to l+\Delta\sigma/k$. }

We work in units of stress in which the modulus $k=1$, and of time in which $\tau_0=1$. We restrict ourselves to homogeneous shear, assuming  that (for example)  no shear bands form either during forward creep or backward recovery. With this simplification, there is no  lengthscale implied by the flow geometry or model. We also set the glass transition temperature $\xg=1$. This amounts to setting the average trap depth in the landscape's prior distribution to unity, thereby setting the typical scale of {\em strain} for local yielding events.  To compare with any experimental data, therefore, all strains reported below would need to be rescaled by the typical strain for local yielding events in the material. 

We set the noise temperature $x=0.3$, deep within the glass phase. Results are presented for a number of SGR elements $M=10^5$, convergence checked against $M=10^6$. The numerical timestep $dt=\alpha/\langle |l|r(E,l)\rangle$ with $\alpha=10^{-5}$, convergence checked against $\alpha=2\times 10^{-6}$. 

Important physical parameters to be explored are then the sample age $\tw$ prior to stress switch-on, the amplitude of the imposed stress $\sig$, and the degree of forward plastic strain  that accumulates before the stress is switched off, $\gamf$. Recall from Sec.~\ref{sec:parameters} that this further defines a switch-off time $\tstop(x,\tw,\sig,\gamf)$.

\begin{figure}[t]
\includegraphics[width=\columnwidth]{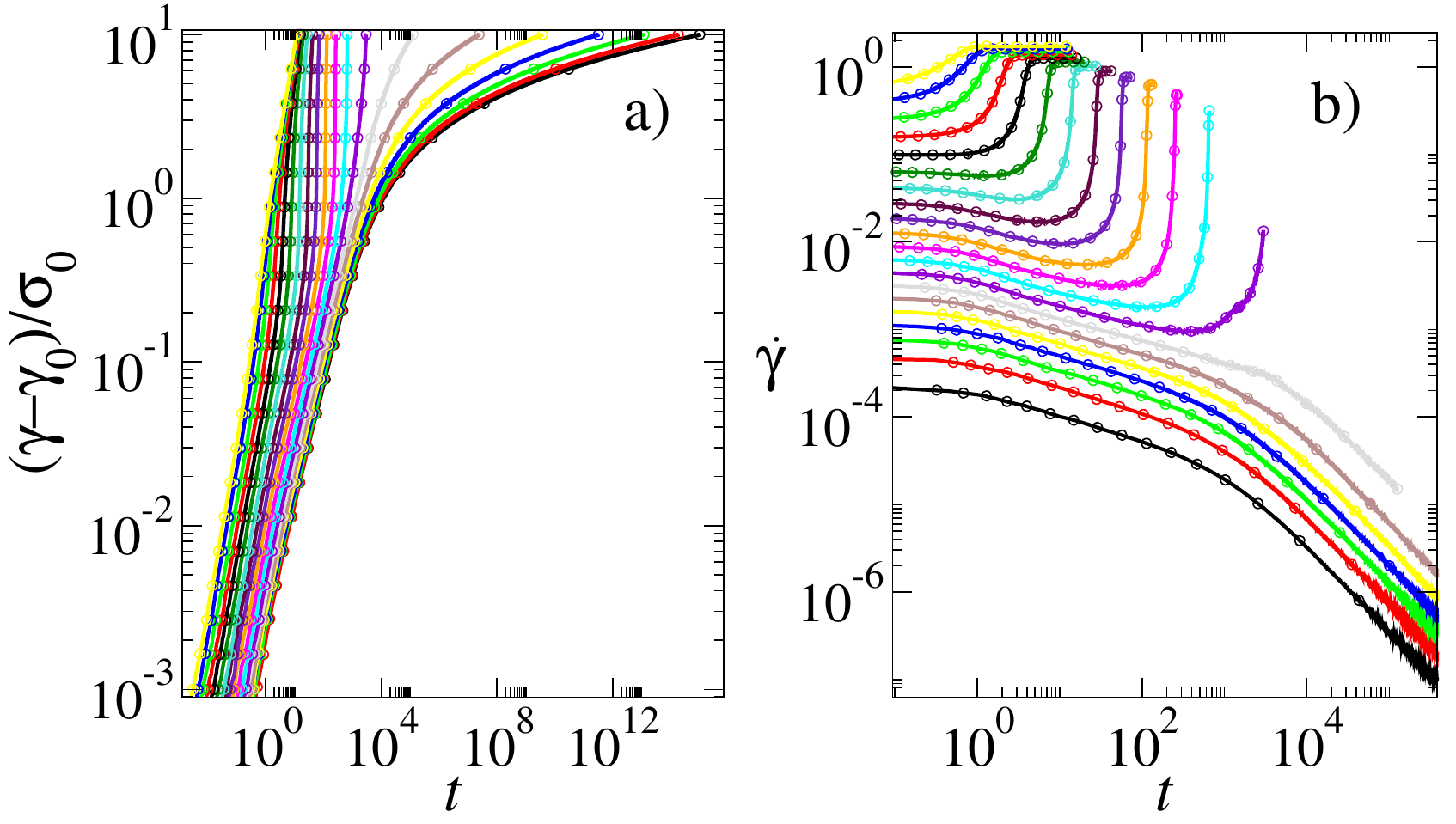}
\caption{Forward strain evolution during stress application. {\bf a)} Plastic strain $\gamma(t)-\gamma_0$  as a function of time $t$ after the initial elastic step  of size $\gamma_0$ that arises at the instant of switch-on $t=0$, normalised by the amplitude of the imposed stress $\sig$. {\bf b)} Corresponding strain rate as a function of time.  Waiting time $\tw=10^3$. Imposed stress $\sig=0.1,0.2\cdots 2.0$ in curves upwards.  Dots are placed at equally logarithmically spaced values of the scaled strain for each imposed stress value in a), and at corresponding times in b). 
}
\label{fig:forward}
\end{figure}

At the instant of stress switch-off, all elements instantaneously strain elastically backwards, giving a global strain change $\Delta\gamma=-\sig$ (in our units). We then measure the subsequent strain evolution as a function of the time $t-\tstop^+$ since switch-off. Our interest in particular will be in the total reverse strain $\gamr$ accumulated in the long time limit, $t-\tstop\to\infty$, beyond that which arose \smfc{in the fast elastic recoil during the reverse step stress}. As noted above, in rare cases we shall find $\gamr < 0$: the \smfc{further straining that arises}  after the initial reverse elastic recoil in fact takes place in the forward direction.

\section{Results}
\label{sec:results}

We now present our results. As a preamble to understanding strain recovery post switch-off, we shall start in Sec.~\ref{sec:forward} by analysing the forward creep and/or yielding and flow that arise while the stress is held imposed. We then present our results for strain recovery in Sec.~\ref{sec:recovery}. 

\subsection{Forward strain during stress application}
\label{sec:forward}

\begin{figure}[t]
\includegraphics[width=0.935\columnwidth]{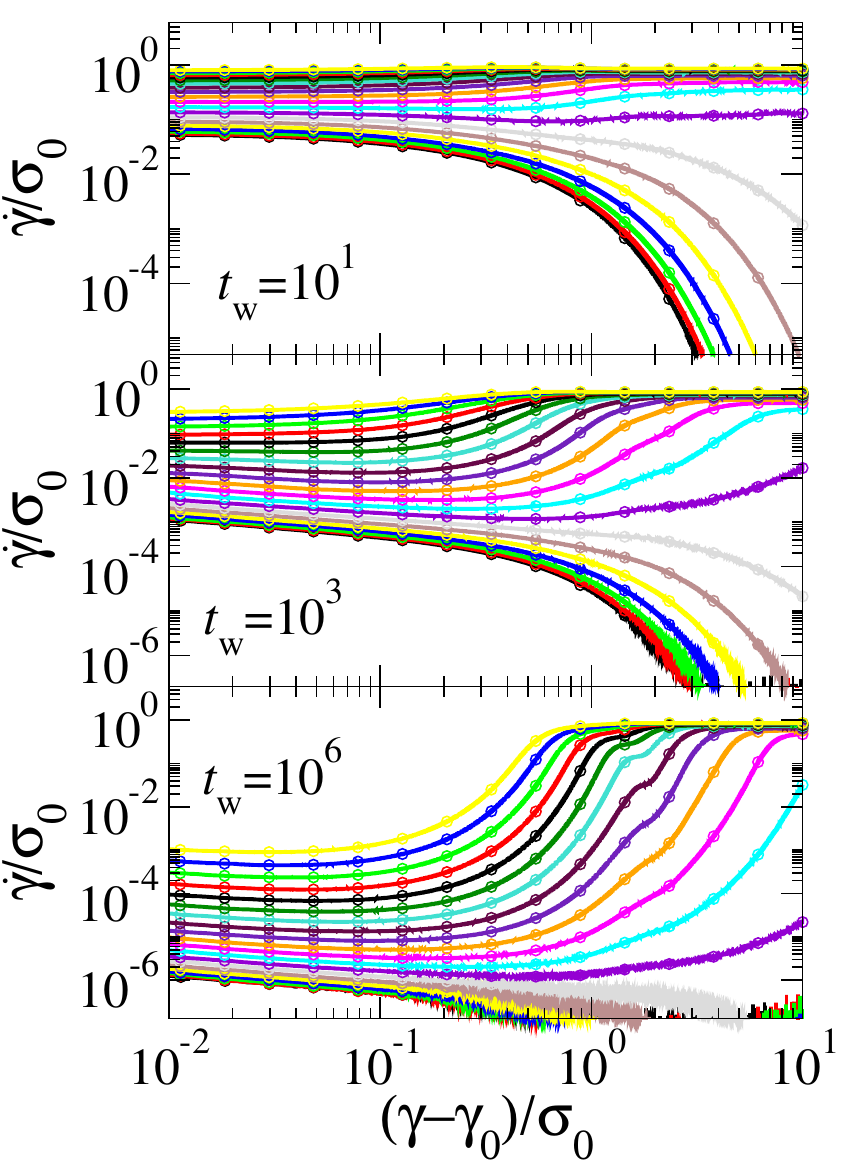}
\caption{Forward strain evolution during stress application. The strain rate is normalised by the imposed stress and parametrically plotted as a function of the plastic strain that arises beyond the initial elastic step strain, also normalised by the imposed stress. Imposed stress $\sig=0.1,0.2\cdots 2.0$ in curves upwards in each panel. Waiting time $\tw=10^1, 10^3$ and $10^6$ indicated in each panel.   Dots show equally logarithmically spaced values of the scaled forward plastic strain  $(\gamma-\gamma_0)/\sig=\gamf/\sig$ for which the subsequent strain recovery after switch-off is reported in Fig.~\ref{fig:recovered}. 
}
\label{fig:forward2}
\end{figure}

Following the application of a step stress of magnitude $\sig$  at time $t=0$, a material's creep response is characterised by the strain curve $\gamma(t)$. As noted above, in the SGR model this shows an initially elastic step strain of magnitude $\gamma_0=\sig/k\;(=\sig$ in our units) at $t=0$, followed by subsequent plastic straining for $t>0$. Fig.~\ref{fig:forward}a) shows this plastic part of the creep, beyond the initial elastic step, normalised by the stress amplitude $\sig$. Results are shown for a fixed sample age $\tw$, and several imposed stresses in curves upwards, \smf{spanning values from below the SGR model's dynamic yield stress $\sigmay$ to above it. (At the noise temperature $x=0.3$ considered in this work, the yield stress $\sigmay=0.758$.)} The corresponding strain rate curves $\gdot(t)$ are shown in panel b).

For imposed stresses $\sig<\sigmay$, the strain rate decreases perpetually as a  power law $\gdot\sim t^{-\alpha}$. The exponent $0<\alpha<1$ depends on the noise temperature $x$. The strain correspondingly increases sublinearly as $\gamma\sim t^{1-\alpha}$, corresponding to sustained power-law creep. In contrast, for imposed stresses $\sig>\sigmay$, an initial interval of creep (for $\sig\gtrapprox \sigmay$ at least) is interrupted by yielding, in which the strain rate curves  upwards before finally settling to a time-independent value prescribed by the steady state flow curve $\sigmass(\gdot)$, with $\sigmass-\sigmay(x)\sim \gdot^{1-x}$. The time delay before yielding scales as the sample age $\tw$, with a prefactor that diverges as $\sig-\sigmay\to 0^+$~\cite{fielding2000aging}.

\begin{figure}[t]
\includegraphics[width=\columnwidth]{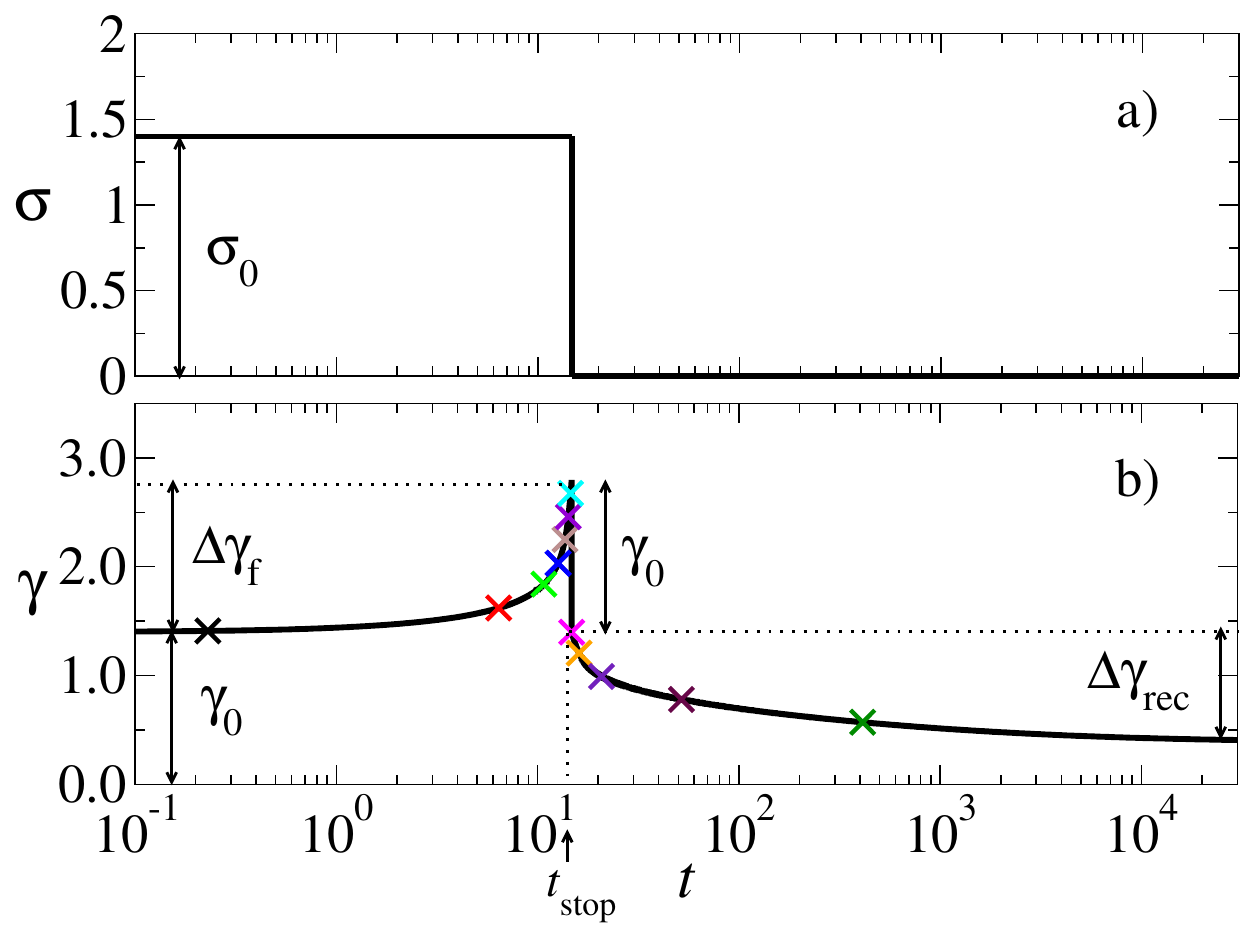}
\caption{{\bf a)} Imposed stress $\sigma$ as a function of time $t$. The sample is prepared at time $t=-\tw$  via a deep temperature quench, then left to age undisturbed for a time $\tw$ (not shown). At time $t=0$ a stress of amplitude $\sig$ is imposed and held constant until a time $\tstop$ (defined below). The stress is then set to  zero and held at zero for all times $t>\tstop$. 
\newline
{\bf b)} Strain response $\gamma(t)$. At the instant the stress is imposed,  $t=0$, the system shows an instantaneous forward elastic step strain of amplitude $\gamma_0=\sig$ (in our units). The strain then subsequently increases beyond $\gamma_0$ due to plastic yielding events. Once it has done so by an amount $\gamf$, the stress is set back to zero. This defines the switch-off time $\tstop$. At this instant  $t=\tstop$, the system shows an instantaneous backward elastic step strain of amplitude $\gamma_0$. The strain then subsequently further recovers in the negative direction (usually) by an additional amplitude $\gamr$.
\newline
Parameters: $x=0.3$, $\sig=1.4$, $\tw=10^3$, $\gamf=1.4$.
}
\label{fig:basic}
\end{figure}

Fig.~\ref{fig:forward2}b) shows the same data as in Fig.~\ref{fig:forward}, but with the scaled strain rate $\gdot(t)/\sig$ now plotted parametrically as a function of the forward plastic strain $(\gamma(t)-\gamma_0)/\sig$. Accordingly,  time $t$ is not explicitly represented in Fig.~\ref{fig:forward2}. The scaled strain on the abscissa however increases monotonically with increasing $t$. The counterpart data for sample ages $\tw=10^1$ and $10^6$ are shown in Figs.~\ref{fig:forward2}a) and c) respectively. As can be seen, for imposed stresses above the yield stress, $\sig>\sigmay$, yielding typically occurs at a scaled strain $(\gamma-\gamma_0)/\sig=O(1)$, although the value increases slowly with $\tw$ and diverges as $\sig\to \sigmay^+$.

A dot at each of several logarithmically spaced values of the abscissa in Fig.~\ref{fig:forward2} shows the values of scaled forward plastic strain $(\gamma-\gamma_0)/\sig=\gamf/\sig$ at which we shall later, in Fig.~\ref{fig:recovered}, explore the amount of strain  $\gamr$ that is ultimately recovered plastically after the stress is switched off.

\subsection{Strain recovery after stress switch-off}
\label{sec:recovery}

Having mapped out the forward creep and/or yielding and flow that arise while the stress is held imposed, we now explore the strain recovery  after the stress is switched off. Our focus  will be  in particular on the part of this recovery that stems from \smfc{slow ongoing plastic yielding of elements that were left with negative local stresses as a result of the fast initial elastic recoil}. \smf{In Sec.~\ref{sec:interplay} below, we shall consider the interplay of a non-zero timescale for the initial (visco)elastic recoil with the later slow plastic strain recovery.}

\begin{figure}[t]
\includegraphics[width=\columnwidth]{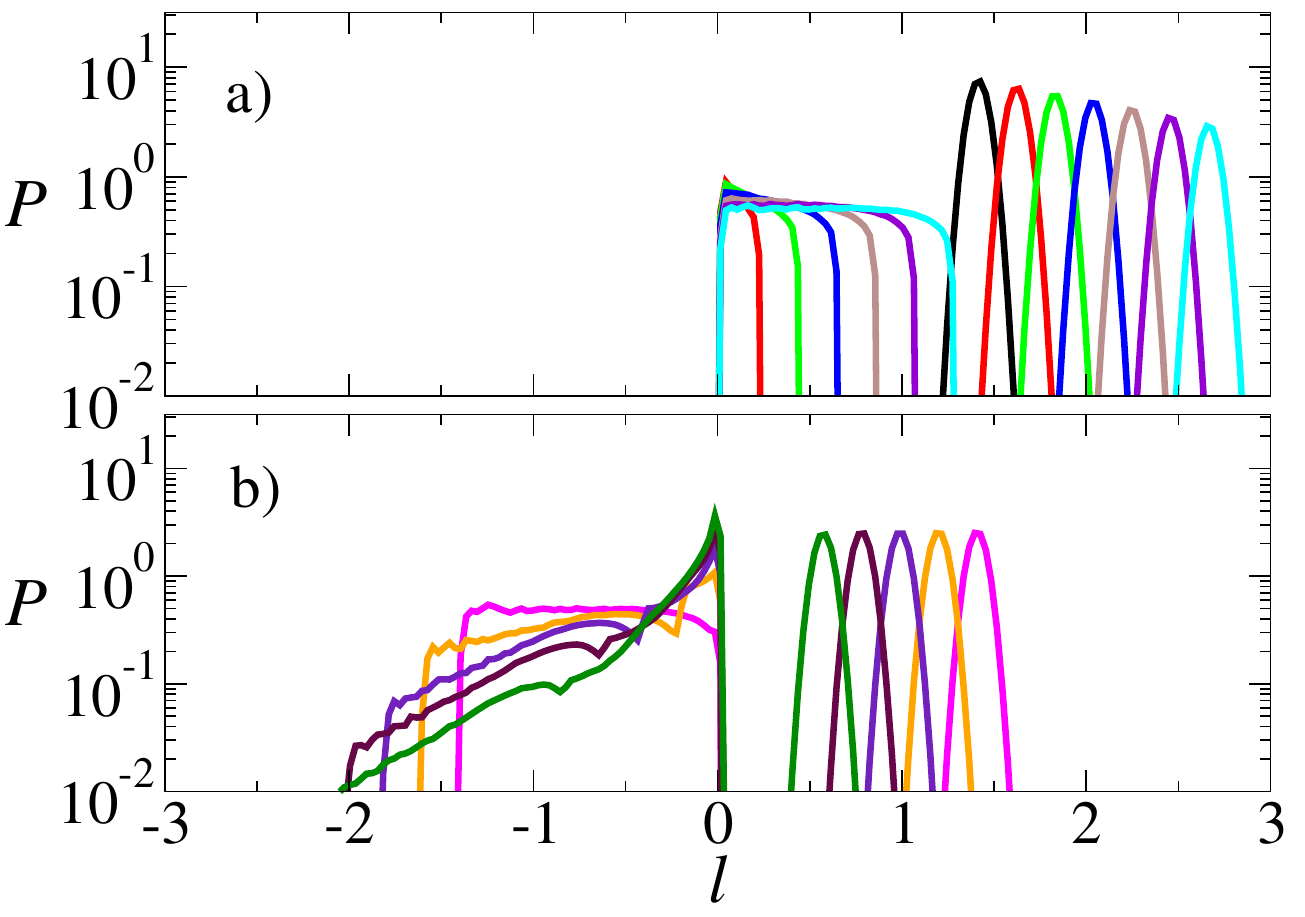}
\caption{Probability distribution $P(l)$ of local elemental strains at several times during the simulation of Fig.~\ref{fig:basic}.  {\bf a)} While the stress is held applied, at times shown by crosses in Fig~\ref{fig:basic}, with time increasing in distributions left to right. {\bf b)} After the stress has been switched off, at times shown by crosses in Fig.~\ref{fig:basic}, with time increasing in distributions right to left. Line colours correspond to cross colours in Fig.~\ref{fig:basic}.
}
\label{fig:distributions}
\end{figure}

The basic physics that we aim to elucidate  is shown in Figs.~\ref{fig:basic} and~\ref{fig:distributions}.  The imposed stress $\sigma$ as a function of time $t$ is shown in the top panel of Fig.~\ref{fig:basic}. The corresponding strain response $\gamma(t)$ is shown in the bottom panel.  As described above, a stress of amplitude $\sig$ is switched on at time $t=0$. (The preparation phase during which the sample is allowed to age undisturbed for a time $\tw$ prior to $t=0$ is not shown.) At the instant of switch-on, the system shows an instantaneous forward elastic step strain of amplitude $\gamma_0=\sig$, in our units in which  $k=1$. (In experimental practice, this forward strain would be rapid but not instantaneous, with a rate limited by inertia or dissipation\smf{; we return to this point in Sec.~\ref{sec:interplay} below}.) The stress is then held constant for some time interval, over which the strain increases further  due to plastic yielding events, this forward straining having been  explored in Sec.~\ref{sec:forward}. Once it has thus increased  by an amount $\gamf$ beyond $\gamma_0$, the stress is switched off. This defines the switch-off time $\tstop$. At the instant of switch-off, $t=\tstop$, the system shows an instantaneous backward elastic step strain, $\Delta\gamma=-\sig=-\gamma_0$. \smf{(We consider in Sec.~\ref{sec:interplay} the effect of a finite timescale for this initial (visco)elastic recoil.)} The stress is then held at zero for all subsequent times $t>\tstop$. As a function of the time interval $t-\tstop^+$, the strain further recovers in the negative direction by an amplitude $\gamr$, beyond the negative elastic step $\Delta\gamma=-\gamma_0$ that occurred at the instant of switch-off. This additional recovery $\gamr$, which has a significant amplitude compared with both $\gamf$ and $\gamma_0$, arises from \smfc{slowly ongoing plastic events, in which elements left with negative local stresses by the initial elastic recoil slowly reset these negative stresses to zero.} 

To understand the physics underlying this strain evolution in more detail, let us consider the master equation governing the time evolution of the joint probability distribution of elemental trap depths $E$ and strains $l$. This was specified in Eqn.~\ref{eqn:master}, which we repeat here for convenience:
\be
\label{eqn:master1}
\dot{P}(E,l,t)+\gdot\frac{\partial P}{\partial l} = -r(E,l)P+Y(t)\rho(E)\delta(l).
\ee
Multiplying across by $l$ and integrating over all $l$ then gives a constitutive equation (albeit not in closed form) for the evolution of the stress, $\sigma$, at imposed strain rate:
\be
\label{eqn:ce1}
\dot{\sigma}(t)=\gdot(t)-\langle l\, r(E,l)\rangle_{P(E,l,t)}.
\ee
In this equation, the first term on the right hand side stems from the elastic loading term that describes the advection of elemental strains while elements remain in their traps. The second term stems from stress relaxation due to local plastic yielding events, as elements hop between traps. The evolution of the strain rate at an imposed stress is then of course simply:
\be
\label{eqn:ce2}
\gdot(t)=\dot{\sigma}(t)+\langle l\, r(E,l)\rangle_{P(E,l,t)}.
\ee

Consider the predictions of this equation for the imposed-stress protocol of interest here. Integrating across the positive (resp. negative) step in stress at $t=0$ (resp. $t=\tstop$) gives an elastic step strain of the same amplitude and sign as the step stress (in our units in which $k=1$), as indeed discussed above and seen numerically in Fig.~\ref{fig:basic}. 

In contrast, during any interval in which the stress is constant, $\dot{\sigma}=0$, either \smf{after switch-on} while the stress of amplitude $\sigma=\sigma_0$ is held imposed, or after switch-off when $\sigma=0$, the strain rate is given by
\be
\label{eqn:ce3}
\gdot(t)=\langle l\, r(E,l)\rangle_{P(E,l,t)}.
\ee
\smf{This tells us that any forward straining $\gamf$ that occurs after the initial elastic loading during switch-on but before switch-off is determined purely by the relaxation of local stresses due to local plastic yielding events. Importantly, it also tells us that} \smfc{{\em any strain recovery $\gamr$ that occurs after the initial elastic recoil post switch-off is likewise determined by the slow plastic yielding of elements left with negative local stress values by that initial recoil.}}  Indeed, as these local plastic events occur and elements hop between traps, a global straining must arise in order to advect elements within their traps and exactly  counterbalance any such plastic stress loss (or gain), keeping the overall stress constant. \smfc{Accordingly, the initial fast elastic recoil creates a state out of which the slow reverse plastic events later emerge, and we return in Sec.~\ref{sec:interplay} below to consider in more detail the interplay between these two different stages of strain recovery.}

In order fully to understand the forward plastic straining during stress imposition and the  strain recovery afterwards, therefore, we must consider in more detail the evolution of the probability distribution of local strains,  and the way  this determines the strain rate $\gdot(t)$ via Eqn.~\ref{eqn:ce3}. Accordingly, we plot this distribution $P(l,t)$ in Fig.~\ref{fig:distributions}  (having first integrated out the trap energy depth variable $E$). We do so at several times $t$ while the stress is held applied (top panel in Fig.~\ref{fig:distributions}), and at several times after the stress is switched off (bottom panel). The time at which any distribution is plotted in Fig.~\ref{fig:distributions} is indicated by a cross of corresponding colour in Fig.~\ref{fig:basic}. 

The form of these distributions in Fig.~\ref{fig:distributions} can be understood as follows. At the end of the sample preparation phase, immediately before the stress is applied, the distribution of local strains is (by assumption) a Gaussian of narrow width $l_0$, centred at $l=0$. (As noted above, our results in all figures except~\ref{fig:basic} and~\ref{fig:distributions} have $l_0=0$, corresponding to a delta function.) At the instant the stress is switched on, this distribution shifts via the elastic advection term, without changing shape, to be now centred at $l=+\sig=+\gamma_0$. This distribution is shown by the black curve in Fig.~\ref{fig:distributions}. Over subsequent times while the stress is held imposed, plastic yielding events arise: recall that an element in a trap of depth $E$ with local strain $l$ has a hopping rate $r(E,l)$ given by Eqn.~\ref{eqn:rate}. When any such plastic event occurs, the corresponding element  resets its local strain $l\to 0$. This slightly depletes the original spike in the  distribution centred at $l=\sig$, and leads to the development of a secondary lobe of probability around $l=0$.

Were the strain held constant, with zero strain rate $\gdot=0$, the macroscopic consequence of these plastic yielding events would be a  relaxation of the ensemble average stress at rate  $\dot{\sigma}=\langle l\, r(E,l)\rangle_{P(t)}$. Conversely, to maintain the stress constant, $\sigma=+\sig$ with $\dot{\sigma}=0$, requires a forward strain of corresponding rate $\gdot=\langle l\, r(E,l)\rangle_{P(t)}$, ensuring sufficient forward advection of elements within their traps exactly to counterbalance this plastic relaxation. This is the origin of the forward plastic straining seen for times $0<t<\tstop$ in Fig.~\ref{fig:basic}. Its consequence is to advect forward at rate $\dot{l}=\gdot$ both the spike in probability centred (originally) at $l= \sig=\gamma_0$ (as it simultaneously depletes due to the plastic relaxation events just described), as well as the newly developing lobe of probability near the origin $l\approx 0$. In this way, the distribution evolves over time from the black to cyan curve in the top panel of Fig.~\ref{fig:distributions}.

At the instant the stress is switched off, \smf{a fast elastic recoil occurs, in which} the distribution of local strains instantaneously shifts by $l\to l -\gamma_0$, with $\gamma_0=\sig$, without changing shape, as each element elastically reduces its strain within its trap. Accordingly, the rightmost (cyan) curve in the top panel of Fig.~\ref{fig:distributions}, which obtained at $t=t_{\rm stop}^-$,  shifts to become the rightmost (magenta) curve in the bottom panel at $t=t_{\rm stop}^+$.

Importantly, this distribution shown in magenta just after switch-off now has some weight for strains $-\gamma_0<l<-\gamma_0+\gamf$, corresponding to those elements that had plastically yielded and reset their strains to zero while the stress was held applied, and then shifted elastically by $l\to l-\gamma_0$ at switch-off. {\em These elements then themselves plastically yield over the subsequent time $t-\tstop$ since switch-off, resetting their strains from that negative strain interval $-\gamma_0<l<-\gamma_0+\gamf$  to $l=0$.} Accordingly, weight is progressively lost from the leftmost part of this probability lobe at negative $l$, with probability instead accumulating nearer $l=0$, in moving from the magenta to green  curves in the bottom panel of Fig.~\ref{fig:distributions}. 

If the strain were held constant, the macroscopic effect of these plastic yielding events from negative to zero $l$ would be a {\em positive} growth in stress, via Eqn.~\ref{eqn:ce3}, as elements hop  from negative to zero local stress. Conversely, under the conditions of constant stress that pertain after stress switch off, $\sigma=0$ and $\dot{\sigma}=0$, a corresponding {\em negative strain rate} is required to advect elements in the negative direction in their traps, in order to counterbalance this positive plastic stress change. Accordingly, as weight is lost from the negative $l$  part of the distribution in moving from the magenta to green curves, these curves also simultaneously advect leftwards.  \smfc{{\em We therefore now understand the slow ongoing strain recovery that arises after the initial elastic recoil post switch-off   to stem from the reverse local plastic yielding of elements left with negative stress values by that initial recoil.} We further recognise these elements to be those that had previously yielded plastically in the forward direction while the material was under load. In this way, counter-intuitively, stress relaxation that occurs plastically while a material is under load can still in fact contribute to strain recovery after the load is removed. }

\begin{figure}[t]
\includegraphics[width=0.85\columnwidth]{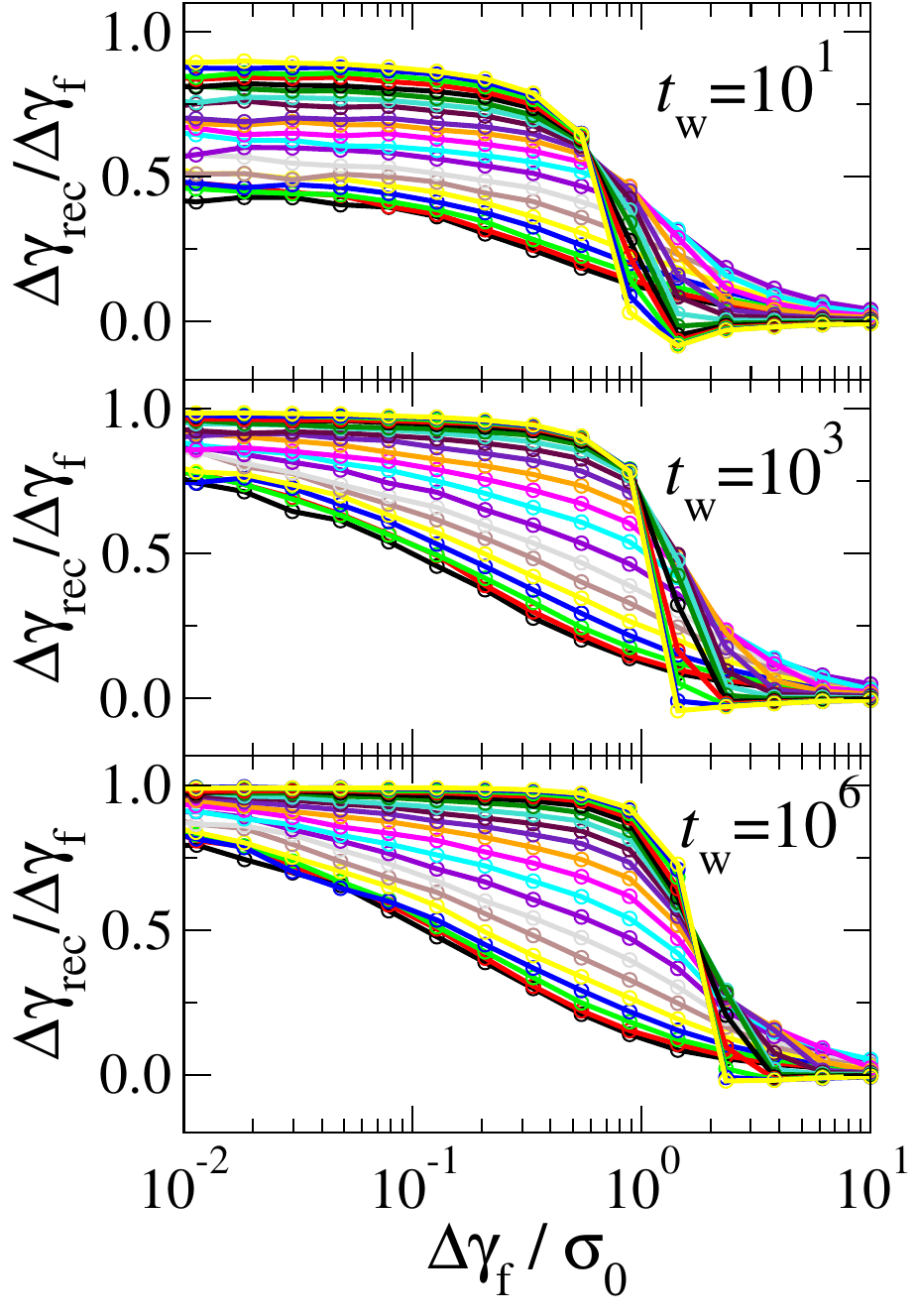}
\caption{Total strain  plastically recovered in the reverse direction at  long times  $t-\tstop\to\infty$ after switch-off, $\gamr$, scaled  by the forward plastic strain $\gamf$ that arose while the stress was held imposed.  Recall that a value $\gamr/\gamf=1$ would indicate full strain recovery. This normalised recovery is plotted as a function of the forward  plastic strain that arose while the stress was imposed, normalised by the imposed stress. Imposed stress $\sig=0.1,0.2\cdots 2.0$ in curves upward in each panel. Waiting time $\tw=10^1, 10^3$ and $10^6$ indicated in each panel.   Dots indicate equally logarithmically spaced values of the scaled forward plastic strain prior to switch-off,  $(\gamma-\gamma_0)/\sig=\gamf/\sig$, corresponding to the dots marked on the differentiated creep curves in Fig.~\ref{fig:forward2}.
}
\label{fig:recovered}
\end{figure}

\begin{figure}[t]
\includegraphics[width=\columnwidth]{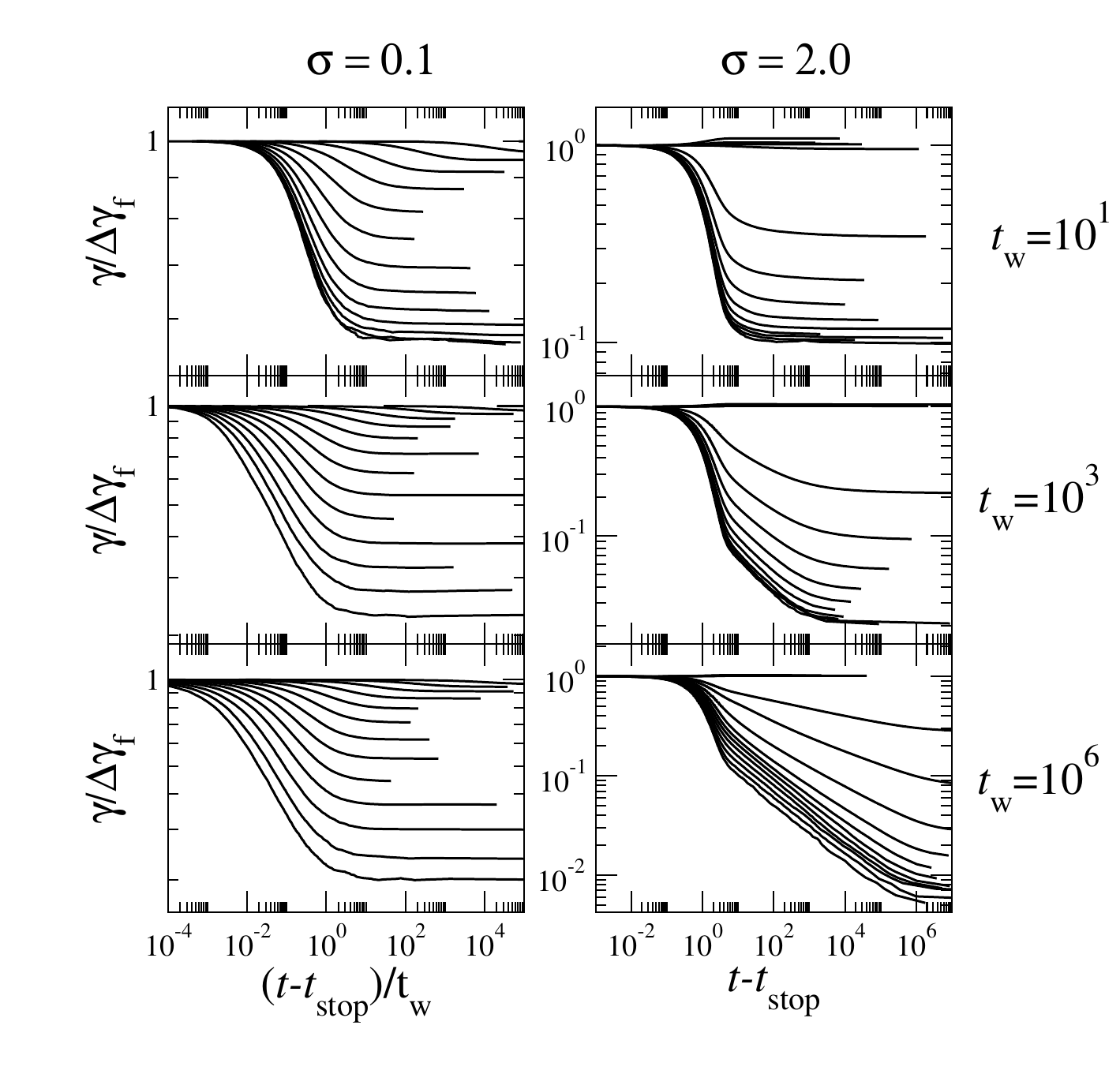}
\caption{Strain recovery as a function of the time since stress switch-off.  The  strain $\gamma$ has been scaled by the forward plastic strain that arose while the stress was held imposed, $\gamf$. The scaled strain $\gamma/\gamf$ accordingly starts at unity at $t=\tstop^+$ (the forward and reverse elastic step strains at switch-on and switch-off having exactly cancelled each other). A full decay of $\gamma/\gamf\to 0$  as $t-\tstop\to\infty$ would indicate full strain recovery. Imposed stress prior to switch-off $\sig = 0.1$ and $2.0$ in left and right panel columns respectively. Sample age prior to stress switch-on $t_w=10^1, 10^3, 10^6$ in panels rows from top to bottom. Plastic forward strain accumulated prior to switch-off  $\gamf = 0.001\sig\times\left(10.0/0.001\right)^{n/19}$ with $n=5,6\cdots$ in curves upwards in each panel.}
\label{fig:recover_vs_time}
\end{figure}

\smf{With this basic physics in mind}, we shall now explore the dependence of the phenomenon on each of the three control parameters in our study: the amplitude of imposed stress $\sig$, the degree to which the sample  plastically strains forward while the stress is held imposed $\gamf$, and the age of the sample $\tw$ before stress switch-on.  We shall consider first in Fig.~\ref{fig:recovered} the total  strain $\gamr$ recovered plastically at long times $t-\tstop\to \infty$, before examining the time-dependence of the recovery process in Fig.~\ref{fig:recover_vs_time}. Recall that $\gamf$ and $\gamr$ are indicated on the left and hand sides respectively of Fig.~\ref{fig:basic}b): a value $\gamr=0$ corresponds to zero plastic strain recovery (beyond the initial elastic recoil $\Delta\gamma=-\gamma_0$ at the instant of switch-off), whereas a value $\gamr=\gamf$ would indicate full recovery.

Accordingly, we plot in Fig.~\ref{fig:recovered} the scaled plastic strain recovery $\gamr/\gamf$ as a function of the scaled forward plastic strain $\gamf/\sig$ that accumulated while the stress was imposed. Data are shown for imposed stress values $\sig=0.1, 0.2\cdots 2.0$ in curves upwards in each panel, for samples ages $\tw=10^1, 10^3$ and $10^6$ in panels downwards, corresponding to the counterpart forward plastic creep curves in  Fig.~\ref{fig:forward2}. Indeed, dots marked left to right on each creep curve in Fig.~\ref{fig:forward2} show the various scaled forward strain values at switch-off, $(\gamma(t)-\gamma_0)/\sig=\gamf/\sig$, shown as corresponding data point dots on each curve in Fig.~\ref{fig:recovered}. 

To understand the trends observed in Fig.~\ref{fig:recovered} we recall that, immediately prior to stress switch-off, any elements that remain unyielded since the stress was imposed have local strain $l=\sig+\gamf$, as in the rightmost lobe of the cyan distribution in Fig.~\ref{fig:distributions}. (This actually has a slightly smeared distribution of strain values of small width $l_0$. As noted above, this is simply for the purposes of being able to plot the distribution graphically and does not change the basic physics.) We shall call these  ``group I" elements. In contrast, those elements that yielded at least once during the initial forward creep (``group II") have local strain values in the range $0<l<\gamf$, giving the leftmost lobe of that distribution. At the instant of stress switch-off, the whole distribution of shifts to the left, with each  $l\to l-\sig$, \smf{corresponding to a fast elastic recoil}.
Immediately after switch-off, therefore, the unyielded group I elements  have positive local strain $l=\gamf$, as in the (slightly smeared) rightmost lobe of the magenta distribution in Fig.~\ref{fig:distributions}. They furthermore reside in traps of a depth set by the age $\tw$, and with a rate of yielding $r \sim 1/\tw$. In contrast, the yielded group II elements have local strain values in the range $-\sig<l<\gamf-\sig$  immediately post switch-off, as in leftmost lobe of the magenta distribution in Fig.~\ref{fig:distributions}.  And in being freshly yielded, they reside in shallow traps with a rate of yielding $r=O(1)$. The relative proportion of elements in group II compared to those in group I increases with increasing forward strain $\gamf$.

Strain recovery post switch-off then arises from a basic competition between the plastic yielding of elements in group I versus those in group II.  In most parameter regimes, 
the  yielding of group II   outweighs that of group I. When these negatively strained group II elements yield, they reset their local strains to zero. As discussed above, to maintain the stress constant and equal to zero post-switch off,  an exactly counterbalancing {\em negative} global strain - {\em strain recovery} - must arise.  

Indeed, for small forward plastic strain values $\gamf<\sig$, all group II elements have negative local strains post switch-off. In this regime, the larger the imposed stress $\sig$ prior to switch-off, the more negative overall will be the local strains of group II elements post switch-off, requiring more counterbalancing  recovery. This explains the increasing levels of recovery   with increasing imposed stress prior to switch-off at the left of Fig.~\ref{fig:recovered}. The larger the sample age $\tw$, the greater the dominance of group II over group I elements, which are stuck in deep traps with yielding rate $~1/\tw$. Accordingly, the degree of strain recovery increases with increasing $\tw$ in moving from the top to the bottom panels of Fig.~\ref{fig:recovered}, for these small $\gamf/\sig$.

A different trend is however observed for  larger values of the scaled forward strain prior to switch-off, $\gamf/\sig\gtrapprox 1$. In this case, for large imposed stresses, the ultimate plastic strain recovery is actually observed to be {\em negative}: the system strains plastically further {\em forwards} as a function of time $t>0$ since switch-off. The overall strain recovery is therefore non-monotonic, with the initially backward elastic recoil at the instant of switch-off $t=\tstop$ followed by this forward plastic straining for $t>\tstop$, reminiscent of observations of non-monotonic stress relaxation after straining~\cite{hendricks2019nonmonotonic,sudreau2022residual,murphy2020memory,mandal2021memory}. The total recovery is still however negative, with the initial elastic recoil always exceeding any subsequent forward plastic straining, for all parameter regimes we have explored.

This counter-intuitive observation of {\em forward} plastic straining post switch-off can be understood as follows. For these larger values of $\gamf\gtrapprox \sig$, the local strain distribution  $-\sig<l<\gamf-\sig$  of group II elements immediately post switch-off  contains some weight at positive strain $l>0$. Furthermore, group I elements are also strained further forwards in their traps, acquiring larger yielding rates. In consequence, as the time since switch-off elapses, the yielding of elements with positive local strains now outweighs that of elements with negative local strains.  To maintain zero stress post switch-off, this must be exactly counter-balanced by an {\em increase} in strain, giving  {\em negative} strain recovery. This effect is more pronounced in younger samples (small $\tw$) for which group I elements are in relatively shallow traps and yield more rapidly. 

For large $\gamf/\sig$, such significant forward straining takes place before switch off, with significant flow for imposed stresses $\sig>\sigmay$, that the proportional degree of recovery post switch off is small. Accordingly,  $\gamr/\gamf\to 0$ as $\gamf/\sig\to \infty$,  as indeed seen in Fig.~\ref{fig:recovered}.

We show finally in Fig.~\ref{fig:recover_vs_time} the strain recovery as a function of the time $t-\tstop$ since switch-off. The six panels shown correspond to the smallest stress (left panel column) and largest stress (right panel column) and three different waiting times (increasing in panel rows downwards) for which the ultimate recovery in the limit $t-\tstop\to\infty$ is plotted versus the scaled forward strain $\gamf/\sig$ in Fig.~\ref{fig:recovered}. Each panel in Fig.~\ref{fig:recover_vs_time} shows results for the time dependent strain recovery for several different values of  forward strain $\gamf$, increasing logarithmically in curves upwards, with the lowest curve in each panel of Fig.~\ref{fig:recover_vs_time} corresponding to the leftmost datapoint in the counterpart curve of Fig.~\ref{fig:recovered}.

\begin{figure}[t]
\includegraphics[width=\columnwidth]{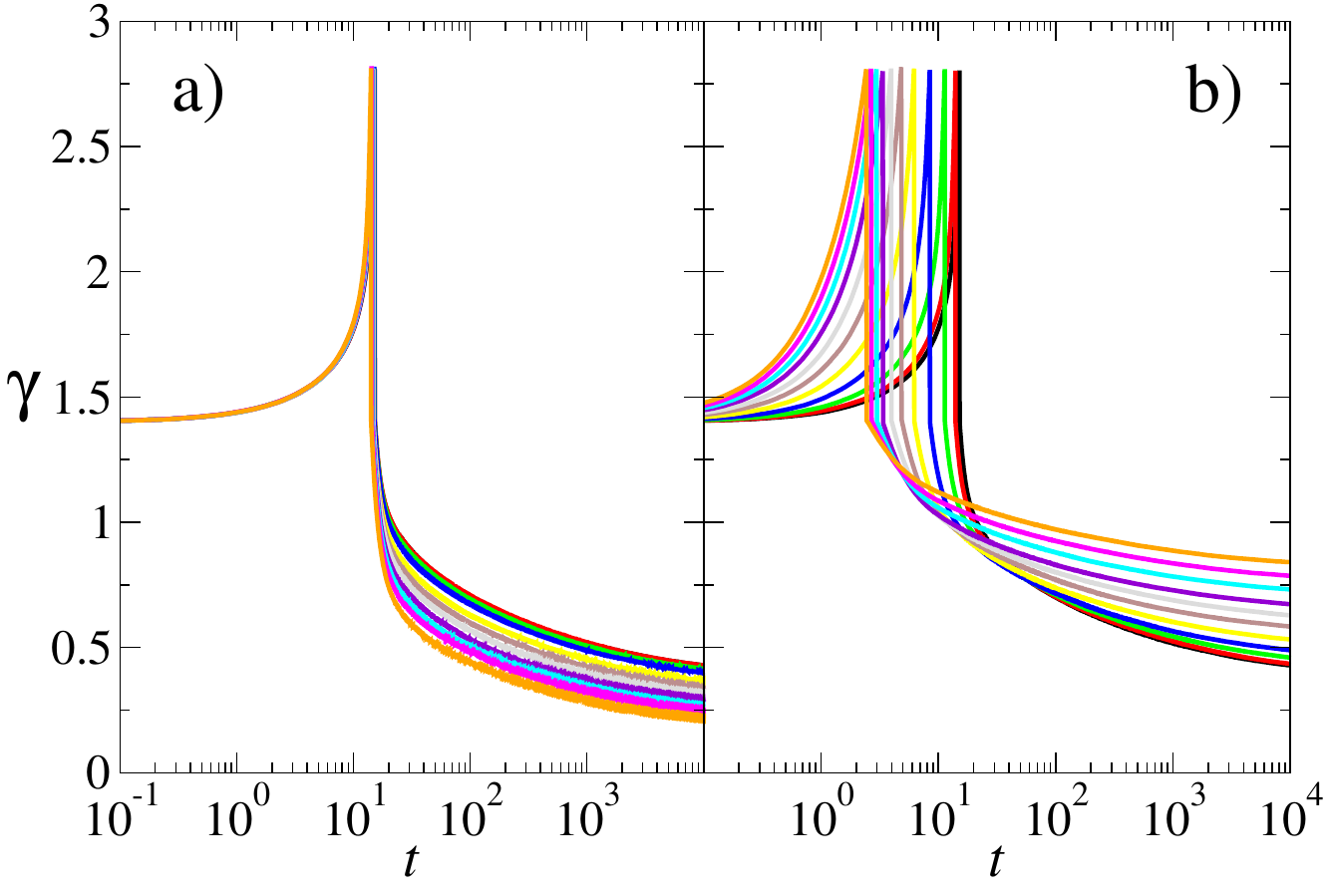}
\caption{\smf{Strain response with non-zero local strain frustration, for parameter values otherwise as in Fig.~\ref{fig:basic}. a) Effect of  non-zero width $l_p$ to the distribution of strains selected by any element immediately post-hop, with $l_p=0.0, 0.1\cdots 1.0$ in curves downwards at the right. b) Effect of a non-zero width $l_0$ to the distribution of strains in the sample at time $t=0$, after sample preparation and before shearing commences, with $l_0=0.0, 0.1\cdots 1.0$ in curves upward at the right.}}
\label{fig:frustration}
\end{figure}

\smf{So far, we have assumed the distribution of local strains present in the sample at time $t=0$, before shearing commences, to be a delta function of zero mean. (An exception can be found in Fig.~\ref{fig:basic}, where we assumed a narrow Gaussian simply to facilitate plotting.) We have likewise assumed the distribution of strains selected by any element immediately post-hop to be a delta function of zero mean. We now consider the more general case in which the distribution before shearing commences is a Gaussian of zero mean and standard deviation $l_0$, and the post-hop distribution is a Gaussian of zero mean and standard deviation $l_p$. (Our results  so far therefore have $l_0=l_p=0$.) The parameter $l_0$ thus now represents the degree of local strain frustration present in the sample before shearing starts, arising from whatever complicated ageing or annealing dynamics took places during sample preparation. Smaller values of $l_0$ correspond to better ordered, more strongly annealed samples. Likewise, $l_p$ represents the frustration experienced by elements in attempting to relax their local strains during local rearrangement events.}

\smfc{In Fig.~\ref{fig:frustration}, we explore the effect of non-zero values of $l_0$ and $l_p$ on the basic phenomenon of recoverable strain arising from reverse plastic events.} \smfb{Shown are results counterpart to those in Fig.~\ref{fig:basic}, but now for a range of $l_p$ values  in panel a) and $l_0$ in ~\ref{fig:basic}b). As can be seen, the basic phenomenon is quite robust to varying these quantities. Indeed, the degree of strain recovery \smfc{from reverse plastic events} increases with increasing post-hop frustration $l_p$, as might be expected intuitively: the left hand lobe of the turquoise distribution (for example) in Fig.~\ref{fig:distributions}a) will now be more smeared, including some weight to the left of the origin. When elastically shifted leftward to form the magenta distribution in b), more weight will be present at more negative $l$ values, resulting in faster yielding of elements with more strongly negative $l$, and more negative global straining ({\it i.e.} more strain recovery) to compensate. And although the effect of plastic strain recovery decreases with increasing initial frustration $l_0$ in panel Fig.~\ref{fig:frustration}b), it nonetheless persists to values of this parameter $O(1)$.}

\section{Interplay of initial (visco)elastic recoil and slow ongoing plasticity}
\label{sec:interplay}

\smf{So far, we have used the SGR model in its original form, considering only the elastoplastic stress arising from  the population of SGR elements. In this version of the model, the initial response to an applied step stress is instantaneous: elements shift elastically within their traps at the instant of the step. Accordingly, in Fig.~\ref{fig:basic} (say) we saw an instantaneous elastic recoil at the time the stress is switched off, followed by the later slow recovery arising from}  \smfc{reverse plastic yielding of elements left with negative local strand by the initial recoil.} \smf{(The instantaneous elastic upward step during switch-on at $t=0$ is not seen on the logarithmic time axis in Fig.~\ref{fig:basic}.) }

\smfc{In practice, however, drag against a background solvent  will render the timescale for this first part of the strain response non-zero, such that the initial recoil is now viscoelastic rather than purely elastic. An obvious question is then to what extent the initial (visco)elastic recoil and later} \smfc{reverse plastic yielding of elements with negative local stresses} \smf{during strain recovery after stress switch-off overlap in time, or whether they occur truly sequentially, with the fast (visco)elastic recoil simply acting to leave elements with negative local stresses that later relax to give reverse} \smfc{plastic events}. \smf{To answer this question, we now consider the total stress $\Sigma=\sigma+\eta\gdot$ to comprise the usual elastoplastic stress $\sigma$ as predicted by  the SGR model in Eqn.~\ref{eqn:sigma}, plus a Newtonian solvent contribution of viscosity $\eta$.}

\begin{figure}[t]
\includegraphics[width=\columnwidth]{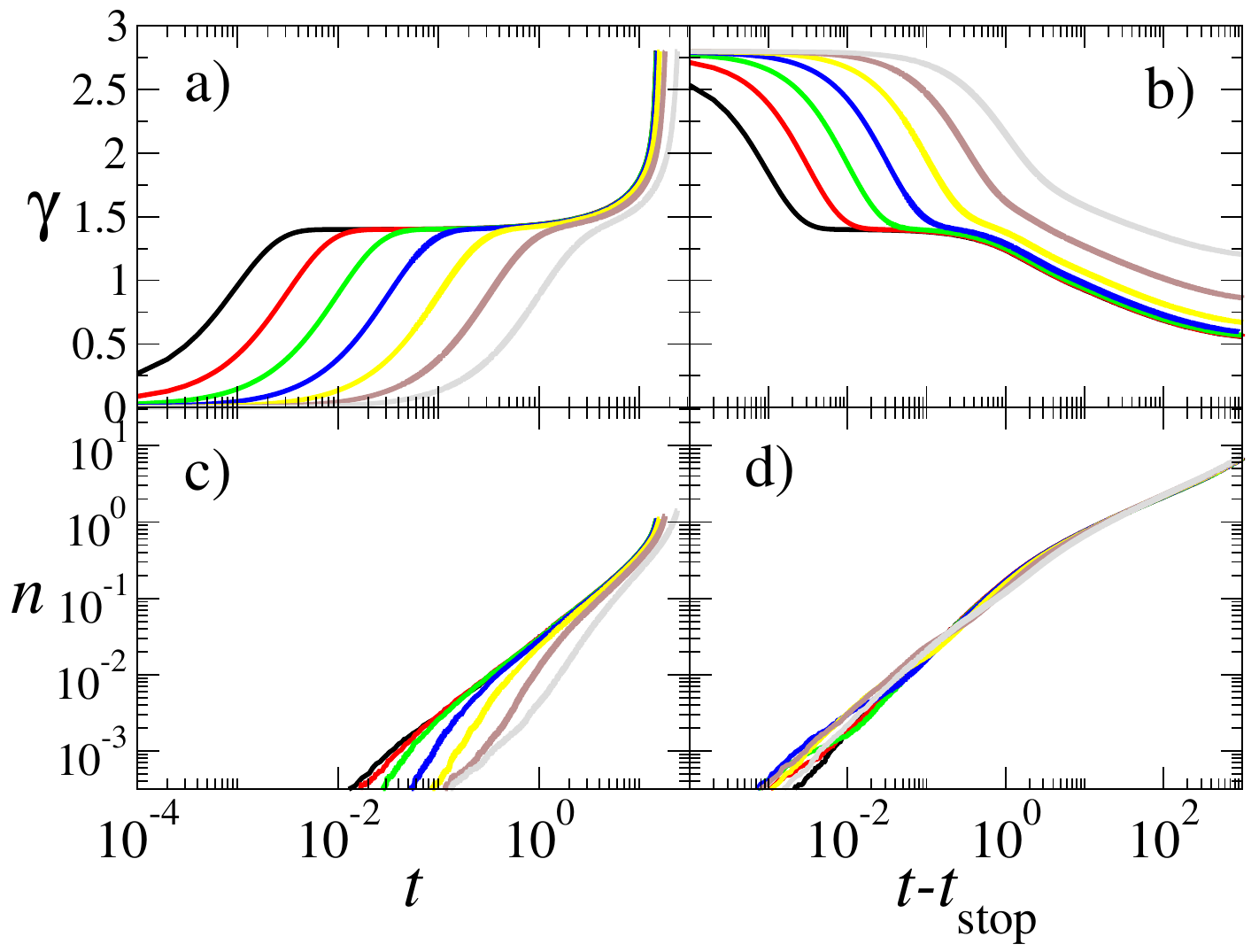}
\caption{\smf{Strain response with non-zero background solvent viscosity $\eta$, for parameter values otherwise as in Fig.~\ref{fig:basic}. a) Forward strain creep while the stress is held imposed. b) Strain recovery after the stress is switched off. c) Number of plastic events per element accumulated while the stress is held imposed. (d) Number of plastic events per element accumulated after the stress is switched off. Viscosity values $\eta=10^{n}$ with $n=-3.0,-2.5\cdots 0.0$ in curves left to right in a), with the same colour code across all sub-panels.}}
\label{fig:viscosity}
\end{figure}

\smf{Fig.~\ref{fig:viscosity} shows results for forward creep and reverse strain recovery with model parameter values otherwise as in Fig.~\ref{fig:basic}, but now with a non-zero solvent viscosity $\eta$. As can be seen, this indeed imposes a finite timescale for the initial rise in strain after stress switch-on (panel a), and for the initial (now visco)elastic strain recoil after stress switch-off (panel b). The timescale for each of these processes  scales as $\eta/k$, tending to zero in the limit $\eta\to 0$. It is in this limit that the initial elastic recoil and later} \smfc{reverse plastic events} \smf{become truly distinct: for finite $\eta$, the two processes merge somewhat.}

\smf{Panels c) and d) show the number of plastic events $n$ accumulated per element on average during forward creep and reverse recovery. Consistent with plastic events being thermally activated (as well as strain-induced) in the SGR model (rather than purely strain-induced, as would be the case in an athermal system), these signals show only modest dependence on the strain (and so on $\eta$). As can be seen, the plastic part of the forward creep and -- importantly to the key message of this work -- of the reverse recovery, arises in each case only once $n$ approaches $O(1)$, denoting significant plastic yielding.}

\smf{We have seen, then, that the initial (visco)elastic recoil post-switch off and the later slow plastic strain recovery will only be truly time-separated in the limit of zero solvent viscosity $\eta\to 0$. In either case (for $\eta$ finite or zero), however, it is clear that these two stages interact, in the sense that the initial (visco)elastic recoil creates the state in which some elements have negative local strain values. It thereby creates the conditions out of which the slower plastic part of the strain recovery can later emerge.}

\section{Implications for constitutive modelling}


\smf{Finally, we consider for the purpose of pedagogical comparison the response of a simplified fluidity model to the same stress protocol as studied for the SGR model in this work. The model is defined as follows. The total stress comprises an elastoplastic contribution $\sigma$ and a Newtonian solvent contribution of viscosity $\eta$:}
\begin{equation}
    \Sigma=\sigma+\eta\gdot.
    \label{eqn:fstress}
\end{equation}
\smf{The elastoplastic stress evolves with dynamics}
\begin{equation}
    \frac{d\sigma}{dt}=G\gdot-\frac{\sigma}{\tau},
\end{equation}
\smf{in which the relaxation timescale $\tau$ has its own dynamics}
\begin{equation}
    \frac{d\tau}{dt}=1-|\gdot|(\tau-\tau_0).
\end{equation}
\smf{Here $G$ is a constant elastic modulus and $\tau_0$ a constant microscopic timescale. (Artificially setting the relaxation time to infinity then gives  Kelvin-Voigt dynamics. Instead artificially setting the viscosity to zero and the relaxation time to a constant  gives a Maxwell model.) The dynamics assumed for the relaxation timescale $\tau$ confers a yield stress, as in SGR, with a flow curve given by $\sigma=G(1+\gdot\tau_0)$. In the absence of any imposed flow the model displays ageing in which the stress relaxation timescale $\tau$ increases linearly in time, also as in SGR.}

\begin{figure}[t]
\includegraphics[width=\columnwidth]{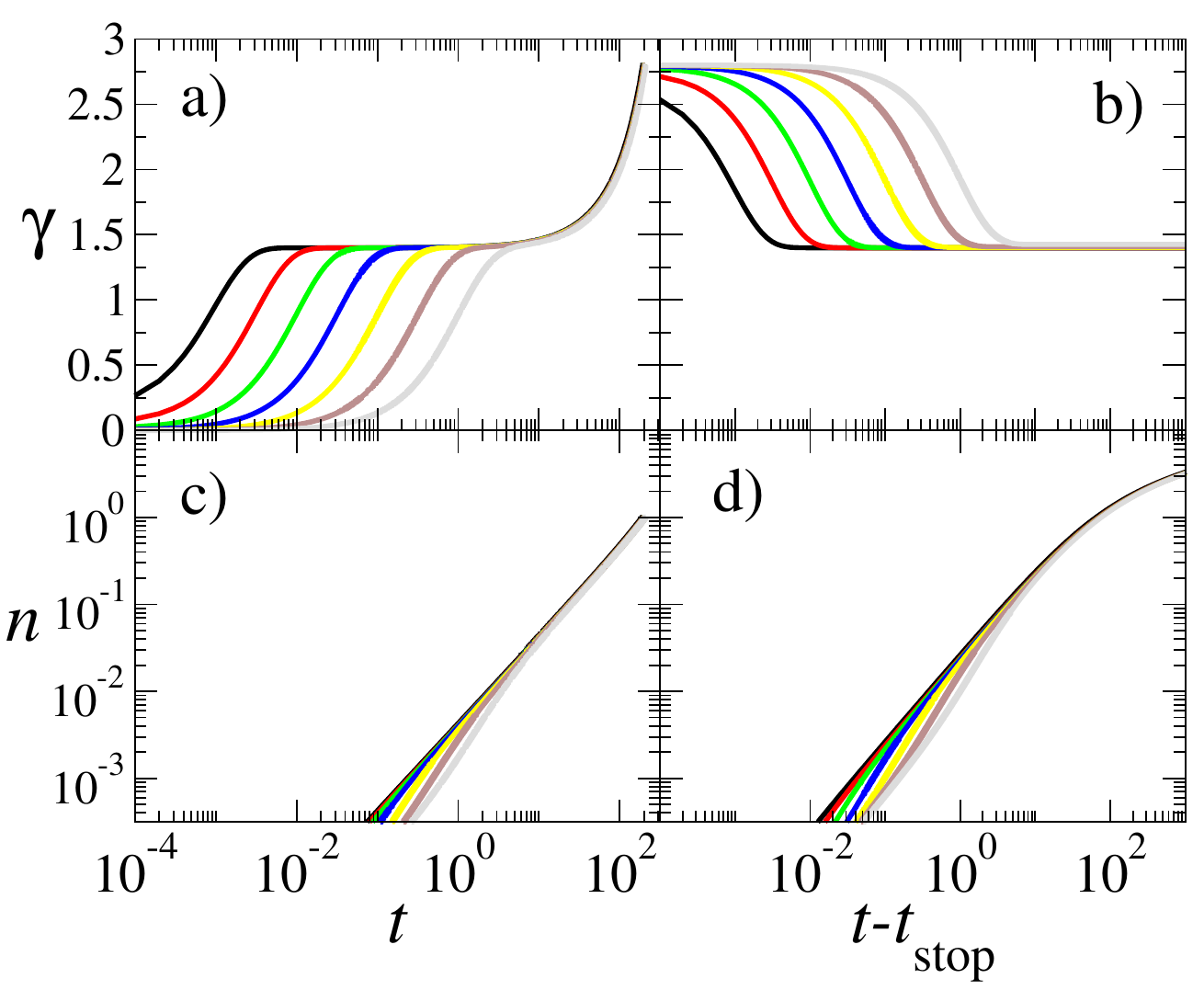}
\caption{\smf{Strain response of simplified fluidity model with non-zero background solvent viscosity $\eta$. a) Forward strain creep while the stress is held imposed. b) Strain recovery after the stress is switched off. c) Integrated rate of plasticity accumulated while the stress is held imposed. (d) Integrated rate of plasticity accumulated after the stress is switched off. Viscosity values $\eta=10^{n}$ with $n=-3.0,-2.5\cdots 0.0$ in curves left to right in a), with the same colour code across all sub-panels. $\Sigma=1.4$, $\tw=10^3$, $\gamf=1.4$.}}
\label{fig:fluidity}
\end{figure}

\smf{The strain response of this simplified model to the stress protocol considered in this work is shown in Fig.~\ref{fig:fluidity}. When the stress is imposed, we see an initial (visco)elastic loading on a timescale $O(\eta/G)$, followed by a slower creep once the integrated rate of plasticity $n=\int dt\, 1/\tau$ approaches $O(1)$, as in SGR. After stress switch-off, we see an initial (visco)elastic recoil on a timescale $O(\eta/G)$, as in SGR. Crucially, however, this simplified model lacks a distribution of local stresses, having just one global averaged stress $\sigma$. Accordingly, it lacks reverse plastic events, and  correspondingly lacks the} \smfc{slow ongoing strain recovery} \smf{predicted by the SGR model: compare panel b) of Fig.~\ref{fig:fluidity} for the fluidity model with the counterpart panel b) of Fig.~\ref{fig:viscosity} for SGR.}

\smfc{This has important implications for constitutive modeling, in showing that the phenomenon studied in this work can be captured only by models that evolve a full distribution of stress values (or multiple moments of such a distribution), rather than a single sample-averaged stress.}

\section{Conclusions}
\label{sec:conclusions}

In this work, we have studied theoretically recoverable and unrecoverable strain in amorphous and soft glassy materials such as dense foams and emulsions, glassy colloidal suspensions, jammed athermal suspensions, granular matter, and metallic and molecular glasses. Within the soft glassy rheology model~\cite{sollich1997rheology,sollich1998rheological}, we have simulated the rheological protocol in which a material of age $\tw$ is subject to the switch-on at time zero of a shear stress of amplitude $\sig$, which is later switched off at some time $\tstop$, and the strain recovery   observed as a function of the subsequent time $t-\tstop^+$. 

We have reported  the counter-intuitive phenomena that a significant component of the strain recovery  post switch-off arises, \smfc{after an initially fast elastic recoil, via the slow reverse plastic yielding of  elements left  with negative local stresses by that recoil}. We have also showed that the time-evolution of strain post  switch-off is not always monotonic, but can be non-monotonic (albeit in relatively rare parameter regimes): the plastic  straining that takes place for times after switch-off $t>\tstop$ does not always  occur in the negative direction, as in the initial elastic recoil at $t=\tstop$, but can sometimes in fact continue to accumulate in the same forward  direction as the originally imposed stress. This provides a potential counterpart, in this imposed-stress protocol, to recent observations of non-monotonic stress relaxation after the switch off of an imposed strain~\cite{hendricks2019nonmonotonic,sudreau2022residual,murphy2020memory,mandal2021memory}.

We  have discussed these observations in terms of the evolution of the SGR model's population of elastoplastic elements. In particular, we have shown that \smf{plastic strain recovery (after the initial fast elastic recoil)} arises via elements that locally yielded  during the initial forward plastic creep, resetting their local strains from positive values to zero as they did so. \smfb{These elements then acquire negative local stresses during the fast elastic recoil, and later yield again post switch-off,  resetting these local stresses from negative values to zero}. It is this plastic yielding from from negative to zero stress values that leads to global \smf{plastic} strain recovery, in order to maintain the global stress of the sample as a whole equal to zero post switch-off. \smfc{In this way, elements that yield plastically in the forward direction while a material is under load can still contribute to strain recovery after load removal.}

\smfc{The phenomenon reported here has important implications for constitutive modelling, in showing that a full distribution of stress values (or multiple moments of such a distribution) is needed to capture it. Any model that evolves just a single average stress will fail to describe it.}

\smfc{The physics described in this work is reminiscent of but distinct from} \smfb{ the concept of {\em reversible plasticity}, which has gained increasing attention in the recent physics literature.  Indeed, a key concept in the rheology of amorphous materials is that of intermittent local plastic  events, sometimes called T1 events, arising at soft spots or shear transformation zones~\cite{falk2011deformation}, coupled by elastic stress propagation~\cite{picard2004elastic}. Within this framework, oscillatory shear experiments on jammed particle rafts identified three different regimes of imposed strain amplitude~\cite{galloway2020quantification,keim2014mechanical,keim2013yielding}. At low amplitudes, the material responds elastically, with no plastic events. At high amplitudes, irreversible plastic events occur within each cycle, with significant cycle-to-cycle stroboscopic  changes in particle positions. More surprisingly, at intermediate strain amplitudes these experiments report {\em reversible local plastic events} within each cycle. Such events dissipate energy, consistent with the notion of plasticity, {\em but exactly reverse in the opposite part of the cycle, with zero stroboscopic cycle-to-cycle change in particle positions}. Reversible plastic events have also been reported in experiments on foams~\cite{lundberg2008reversible}, 
in particle simulations~\cite{lundberg2008reversible,regev2013onset,priezjev2016reversible,szulc2022cooperative,nagasawa2019classification} and in elastoplastic models~\cite{khirallah2021yielding}. Reversible plastic limit cycles have likewise been studied in models of  hysterons~\cite{keim2021multiperiodic} and iterated maps~\cite{mungan2019cyclic}. For a review, see Ref.~\cite{reichhardt2022perspective}. }

\smf{It should be noted, however, that the mean field SGR model as studied here does not provide any spatial resolution of elemental positions. Accordingly, we are not able to make any statement as to whether a plastic event at a specific location later exactly reverses at the same location. Furthermore, because plastic events in SGR are thermally activated as well as being strain-induced, exact reversibility under straining is unlikely, except in the athermal limit $x\to 0$ -- in which limit, however, predictions of sustained slow creep are difficult to obtain.} 

\smf{Therefore, we do not claim a close correspondence between the plasticity that arises in this work {\em in the reverse direction during strain recovery post-switch off} and the truly {\em reversible plasticity in which individual elements exactly retrace their trajectories}, as seen under oscillatory shear in athermal systems. Instead, we make only the weaker statement that some elements  that yield during the initial forward creep later do so in reverse during recovery. For this reason, we have been careful to use the weaker term} \smfc{{\em reverse plastic events} rather than the the term {\em reversible plasticity}}, \smfb{as coined in the literature concerning oscillatory shear in athermal systems.} \smfc{Indeed, we use the term ``reverse'' in this work to mean ``in the opposite direction", as distinct from ``reversible'' meaning ``an event capable of being reversed''.} \smf{Future work should study creep and recovery in spatially resolved athermal models of amorphous materials, to establish whether there is a stronger link to reversible plasticity, with its precise connotations of exact localised reversibility.}

Related to this, we have assumed throughout that the shear remains macroscopically homogeneous across the sample, disallowing any possibility of shear band formation. However, it has been predicted that shear bands will tend to form in an imposed stress protocol at the end of any initial creep regime, as the strain rate dramatically increases and the sample yields~\cite{mooney1936rheology}. This is likely to be an important effect for large values of $\gamf$ in particular. Future studies should consider extending this work to allow for the formation of shear bands.

{\it Acknowledgements ---}  The authors thank Mike Cates, Andrew Clarke and Gareth McKinley for comments on an early draft of the manuscript, and for helpful discussions. We also thank SLB (Schlumberger Cambridge Research Ltd.) for support. This project has received funding from the European Research Council (ERC) under the European Union's Horizon 2020 research and innovation programme (grant agreement No. 885146). \smfc{We thank the anonymous reviewers for their insightful comments that improved the clarity of our descriptions.}

{\it Author declarations ---}  The authors have no conflicts to disclose.

\section{References}


%

\end{document}